\documentclass[aps,prl,twocolumn,superscriptaddress,citeautoscript,showkeys]{revtex4-2}

\pdfoutput=1
\usepackage{graphicx}
\usepackage{amsmath, amsfonts}
\usepackage{amssymb}
\usepackage[colorlinks=true,linkcolor=blue,citecolor=blue]{hyperref}
\usepackage[utf8]{inputenc}
\usepackage{color}

\begin{document}

\title{Picosecond Femtojoule Resistive Switching in Nanoscale VO$_{2}$ Memristors}

\author{Sebastian Werner Schmid}
\affiliation{Department of Physics, Institute of Physics, Budapest University of Technology and Economics, M\H{u}egyetem rkp. 3, H-1111 Budapest, Hungary}
\affiliation{Experimental Physics V, Center for Electronic Correlations and Magnetism, University of Augsburg, Augsburg 86159, Germany}

\author{L\'{a}szl\'{o} P\'{o}sa}
\affiliation{Department of Physics, Institute of Physics, Budapest University of Technology and Economics, M\H{u}egyetem rkp. 3, H-1111 Budapest, Hungary}
\affiliation{Institute of Technical Physics and Materials Science, HUN-REN Centre for Energy Research, Konkoly-Thege M. \'{u}t 29-33, 1121 Budapest, Hungary}

\author{T\'{i}mea N\'{o}ra T\"{o}r\"{o}k}
\affiliation{Department of Physics, Institute of Physics, Budapest University of Technology and Economics, M\H{u}egyetem rkp. 3, H-1111 Budapest, Hungary}
\affiliation{Institute of Technical Physics and Materials Science, HUN-REN Centre for Energy Research, Konkoly-Thege M. \'{u}t 29-33, 1121 Budapest, Hungary}

\author{Botond S\'{a}nta}
\affiliation{Department of Physics, Institute of Physics, Budapest University of Technology and Economics, M\H{u}egyetem rkp. 3, H-1111 Budapest, Hungary}
\affiliation{HUN-REN-BME Condensed Matter Research Group, M\H{u}egyetem rkp. 3, H-1111 Budapest, Hungary}

\author{Zsigmond Pollner}
\affiliation{Department of Physics, Institute of Physics, Budapest University of Technology and Economics, M\H{u}egyetem rkp. 3, H-1111 Budapest, Hungary}

\author{Gy\"{o}rgy Moln\'{a}r}
\affiliation{Institute of Technical Physics and Materials Science, HUN-REN Centre for Energy Research, Konkoly-Thege M. \'{u}t 29-33, 1121 Budapest, Hungary}

\author{Yannik Horst}
\affiliation{Institute of Electromagnetic Fields, ETH Zurich, Gloriastrasse 35, 8092 Z\"{u}rich, Switzerland}

\author{J\'{a}nos Volk}
\affiliation{Institute of Technical Physics and Materials Science, HUN-REN Centre for Energy Research, Konkoly-Thege M. \'{u}t 29-33, 1121 Budapest, Hungary}

\author{Juerg Leuthold}
\affiliation{Institute of Electromagnetic Fields, ETH Zurich, Gloriastrasse 35, 8092 Z\"{u}rich, Switzerland}

\author{Andr\'{a}s Halbritter}
\affiliation{Department of Physics, Institute of Physics, Budapest University of Technology and Economics, M\H{u}egyetem rkp. 3, H-1111 Budapest, Hungary}
\affiliation{HUN-REN-BME Condensed Matter Research Group, M\H{u}egyetem rkp. 3, H-1111 Budapest, Hungary}

\author{Mikl\'{o}s Csontos}
\affiliation{Institute of Electromagnetic Fields, ETH Zurich, Gloriastrasse 35, 8092 Z\"{u}rich, Switzerland}


\begin{abstract}

Beyond-Moore computing technologies are expected to provide a sustainable alternative to the von Neumann approach not only due to their down-scaling potential but also via exploiting device-level functional complexity at the lowest possible energy consumption. The dynamics of the Mott transition in correlated electron oxides, such as vanadium dioxide, has been identified as a rich and reliable source of such functional complexity. However, its full potential in high-speed and low-power operation has been largely unexplored. We fabricated nanoscale VO$_{2}$ devices embedded in a broad-band test circuit to study the speed and energy limitations of their resistive switching operation. Our picosecond time-resolution, real-time resistive switching experiments and numerical simulations demonstrate that tunable low-resistance states can be set by the application of 20~ps long, $<$1.7~V amplitude voltage pulses at 15~ps incubation times and switching energies starting from a few femtojoule. Moreover, we demonstrate that at nanometer-scale device sizes not only the electric field induced insulator-to-metal transition, but also the thermal conduction limited metal-to-insulator transition can take place at timescales of 100's of picoseconds. These orders of magnitude breakthroughs open the route to the design of high-speed and low-power dynamical circuits for a plethora of neuromorphic computing applications from pattern recognition to numerical optimization.

\end{abstract}


\date{\today}
\maketitle

\noindent \textbf{Introduction}\\

Few-component analog circuits exploiting the dynamical complexity of memristors enable increased functional complexity at a minimized footprint. Moreover, they can energy-efficiently replace extensive digital circuits designed to facilitate demanding algorithms \cite{Kumar2022,Zidan2018a}. Meanwhile reconfigurability, a great asset of digital platforms, has been put forward also in the analog domain by the concept of memristive field-programmable analog arrays \cite{Li2022}. In particular, second order dynamical complexity arising from the Mott-type insulator-to-metal transition (IMT) \cite{Kim2004,Liu2018,Ko2008,Kim2010,Wu2011a,Tadjer2017,Posa2023} in NbO$_{2}$ has been utilized to realize relaxation oscillators and chaotic dynamics for accelerating probabilistic optimization in transistor-less circuits \cite{Kumar2017a,Kumar2017b}. Passively coupled VO$_{2}$ Mott-memristors have been used to mimic the fundamental adaptability of the biological nervous system to various input streams and even faithfully reproduce a spectrum of bio-realistic neural response patterns \cite{Yi2018}. The phase information in coupled VO$_{2}$ oscillator networks has been identified as a state variable and exploited in neural network operation \cite{Corti2018,Corti2019,Corti2020,Corti2021}. The discovery of short and longer enduring metallic domains enabled sub-threshold operation, short-term and long-term memory functionalities extending the toolkit for neuromorphic data storage and processing \cite{Valle2019,Nikoo2022}. 

The potential for high-speed and low-energy operation relying on the IMT in VO$_{2}$ based devices has been predicted by theoretical considerations \cite{Liu2018} and simulations \cite{Yi2018}. Experiments carried out in the optical and THz domain \cite{Becker1994,Cavalleri2001,Cavalleri2004,Wegkamp2014,Morrison2014,Tao2016,Liu2012c} have demonstrated that due to its predominantly electronic origin, the IMT can indeed be completed at sub-picosecond time-scales. However, the more scalable electrical domain applications could so far only exploit $\geq$300~ps set and $\gg$1~ns reset switching times at $\geq$100~fJ energy costs \cite{Valle2019,Valle2021,Nikoo2022}.

\begin{figure*}[t!]
     \includegraphics[width=2\columnwidth]{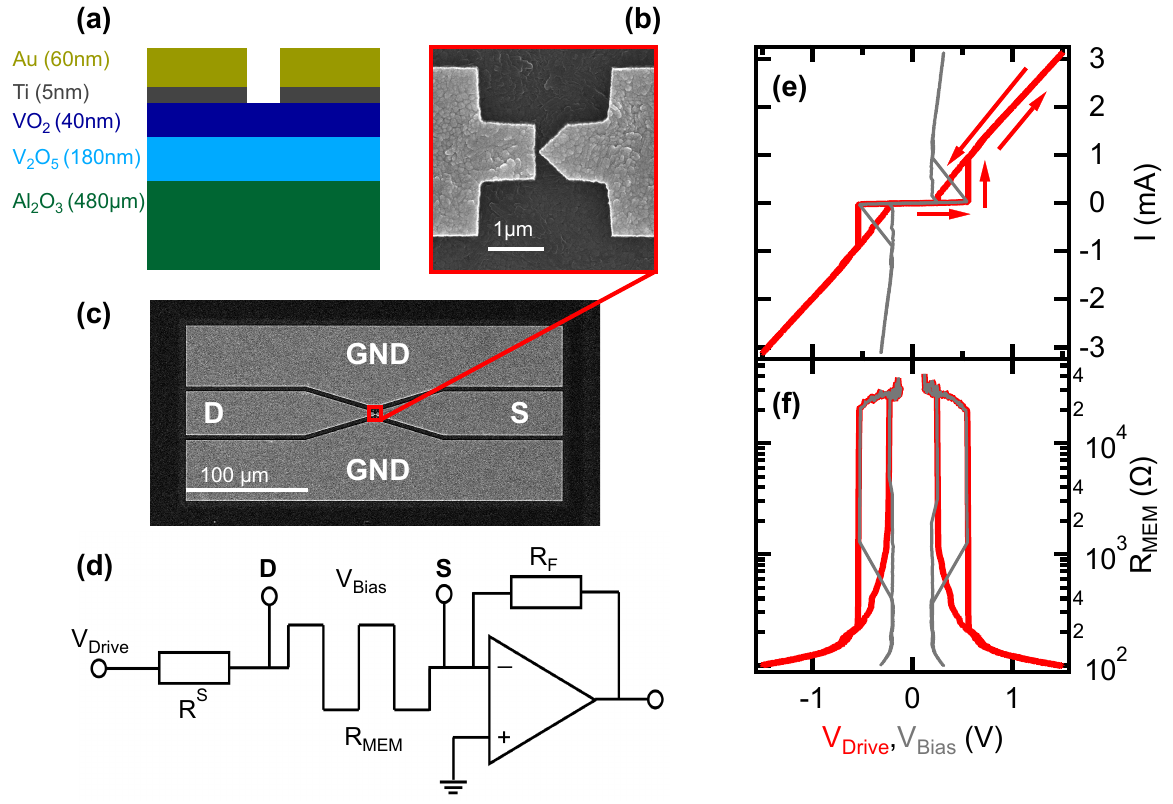}
     \caption{\textbf{Device structure and DC characterization of the resistive switching.} (a) Schematic (not to scale) vertical cross-section of the VO$_{2}$ memristors. (b) Magnified top-view of the planar device. The Au terminals are separated by a $\approx$30~nm gap at the tip of the triangular electrode. (c) Scanning electron microscope image of the co-planar waveguide (CPW) hosting the device for the high-speed resistive switching experiments (S: source, D: drain, GND: ground). The separation of the ground terminals is 4~$\mu$m. (d) The DC $I(V)$ measurement setup includes a series resistor $R ^{\rm S}$, the memristor device and a current amplifier. The $V_{\rm Drive}$ voltage signal is sourced from a data acquisition card whereas the $V_{\rm Bias}$ voltage drop on the memristor is calculated as $V_{\rm Bias}=V_{\rm Drive}-I\cdot R^{\rm S}$. (e-f) Low-frequency ($f_{\rm Drive}=$ 1~Hz) $I(V)$ traces and the calculated resistance, respectively, as a function of $V_{\rm Drive}$ (red) and $V_{\rm Bias}$ (gray). $R^{\rm S}=$ 380~$\Omega$. The red arrows indicate the direction of the hysteresis.}
     \label{fig1}
\end{figure*}

Here we demonstrate that the resistive switching response of VO$_{2}$ memristors to purely electrical stimuli can be as fast as 15~ps for the set transition, in agreement with finite element simulations based on a two-dimensional resistor network model. The evaluation of the current acquired during the applied 20~ps long voltage pulses reveals that the IMT is triggered by the injection of as little Joule heat as a few femtojoule. Furthermore, a dedicated V-shaped electrode arrangement is utilized to focus the electric field and, thus, the device operation to a nanometer-scale active volume. Consequently, the thermal relaxation time to the insulating state of VO$_{2}$ can be greatly reduced to the 100~ps time-scale, enabling low-power dynamical memristor circuits for ultra-fast neuromorphic computing applications.

\noindent \textbf{Results and Discussion}\\

Our report is organized as follows. First, the device structure and the DC characterization of the resistive switching cycles are presented. Next, the transmission spectroscopy of sub-nanosecond voltage pulses is explained. Using this method, we demonstrate the analog tunability of the volatile low resistance states, set switching times down to 15~ps and switching energies in the femtojoule regime. We show that these results can be quantitatively understood in terms of a two-dimensional resistor network of nanometer-scale VO$_{2}$ domains, where the resistance of each domain is determined by the local temperature and electric field. Finally, we use a pump-probe scheme utilizing 20~ps long voltage pulses to monitor the complete recovery of the high resistance state in the sub-nanosecond time-domain.\\

\textbf{Device Structure and DC Characterization.} The individual layer thicknesses of the sample are labeled in the schematic vertical cross-section in Fig.~\ref{fig1}(a). The magnified top view of the planar device is exhibited in Fig.~\ref{fig1}(b). The two, asymmetrically shaped Au electrodes were evaporated on the top of a VO$_{2}$/V$_{2}$O$_{5}$ film, which was created by the thermal oxidation of a V layer evaporated on top of a sapphire substrate \cite{Posa2023,Posa2021}. The planar gap between the two electrodes was around 30~nm. The latter, together with the flat (triangular) shape of the source (drain) electrode, facilitate low-voltage resistive switching in a single, nanometer-scale volume of the underlying VO$_{2}$ layer. Further details on sample fabrication and characterization are provided in the Methods and Experimental Section and in Ref.~\citenum{Posa2023}. Figure~\ref{fig1}(c) shows the electrode layout of the devices. The source (S), drain (D) and the two ground (GND) electrodes made of 50~nm thick Au form a co-planar waveguide (CPW). The CPW ensures the suitable termination of the two-terminal Au/VO$_{2}$/Au devices for the short-duration voltage pulses of the high-speed resistive switching experiments.

\begin{figure}[t!]
     \includegraphics[width=\columnwidth]{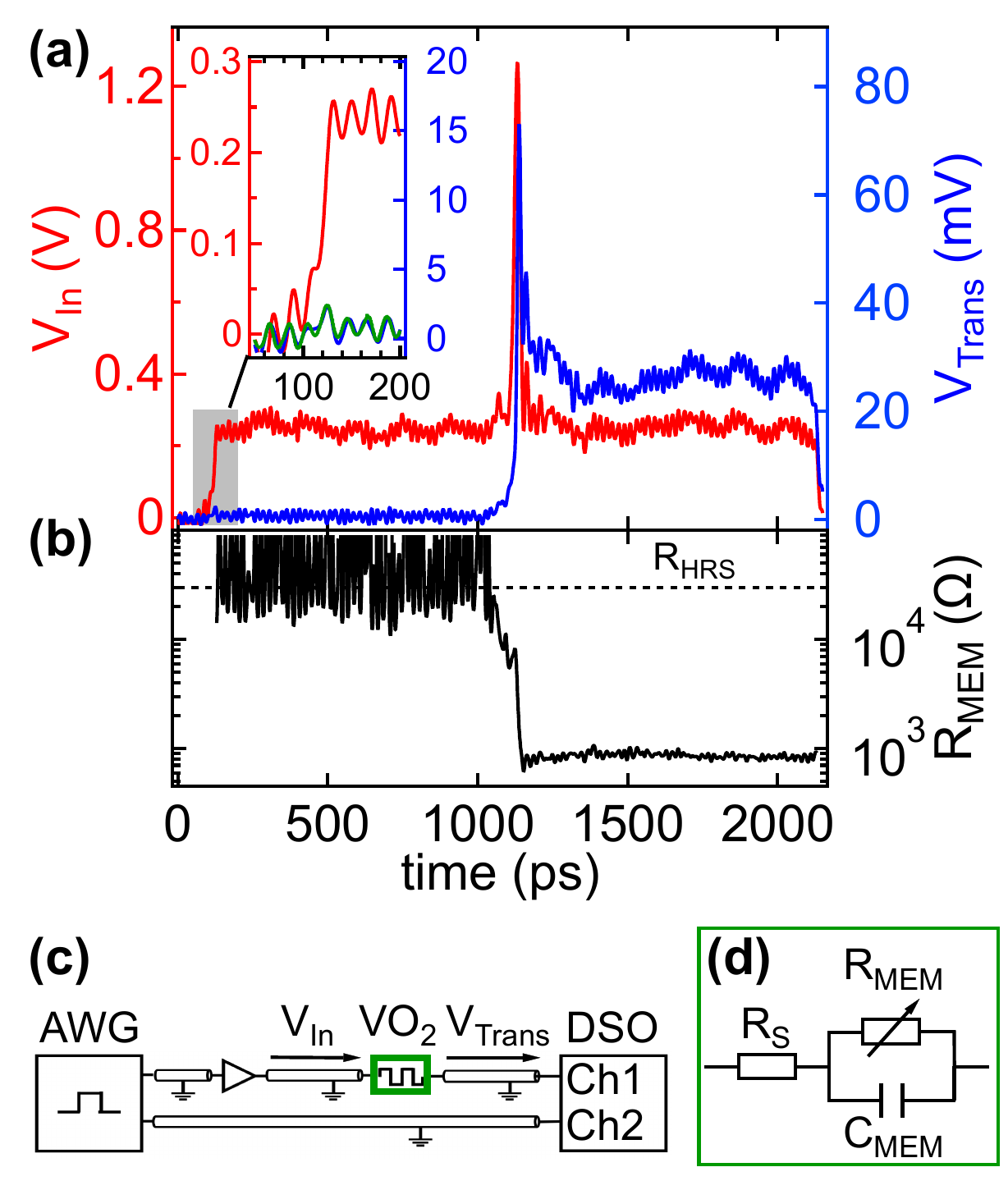}
     \caption{\textbf{Experimental demonstration of picosecond time-scale resistive switching.} (a) Incoming voltage signal $V_{\rm In}$ (left axis, red) consisting of a 2~ns long, low-voltage read-out offset and a higher amplitude, 20~ps FWHM set voltage pulse. The corresponding transmitted voltage $V_{\rm Trans}$ evidences that resistive switching takes place during the set voltage pulse, as shown in blue (right axis). Note the different scales of the vertical axes. The inset magnifies the shaded regime at the onset of the read-out pulse. The negligible capacitive response during the 20~ps rise time of the $V_{\rm In}$ step demonstrates the low, $\approx$2~fF parasitic capacitance of the memristor sample, in agreement with simulations (green), and indicates the dominantly resistive nature of the switching in the investigated resistance regime. (b) The time dependence of the device resistance is calculated from the $V_{\rm In}$ and $V_{\rm Trans}$ traces shown in (a), according to Eq.~\ref{trans.eq}. (c) The schematics of the high-speed setup. The $V_{\rm In}$ amplified voltage pulses are sourced from an arbitrary waveform generator (AWG). The $V_{\rm Trans}$ transmitted voltage is measured by a digital storage oscilloscope (DSO). The DSO is triggered by the second, non-amplified output channel of the AWG. (d) The equivalent circuit of the memristor device. $Z_{\rm MEM}$ consists of a series resistance $R_{\rm S}$ taking the lead and probe contact resistances into account, the device resistance $R_{\rm MEM}$ and the parasitic device capacitance $C_{\rm MEM}$.}
     \label{fig2}
\end{figure}

The DC characterization of the current-voltage [$I(V)$] traces were carried out in a setup consisting of a series resistor, the memristor device, a current amplifier and a data acquisition card which was also utilized as a programmable voltage source, as illustrated in Fig.~\ref{fig1}(d) and explained in detail in the Methods and Experimental Section. In our nomenclature the $V_{\rm Drive}$ voltage is applied on the device and the current limiting series resistor $R^{\rm S}$ whereas the voltage drop on the memristor device only is denoted as $V_{\rm Bias}$. Positive voltage corresponds to higher potential on the drain electrode.

A representative, hysteretic $I(V)$ trace exhibiting unipolar, volatile resistive switching is shown in Fig.~\ref{fig1}(e) both as a function of $V_{\rm Drive}$ (red) and $V_{\rm Bias}$ (gray). The resistance, calculated as $V_{\rm Bias}/I$ is plotted in Fig.~\ref{fig1}(f). Resistive switching reproducibly occurs between typical high resistance states (HRS) of $R_{\rm HRS}\approx$ 30~k$\Omega$ and tunable low resistance states (LRS) in the 10$^{2}$\,--\,10$^{3}$ regime at set (reset) voltages around 550~mV (250~mV). The actual $R_{\rm LRS}$ value can be fine-tuned by the choice of $R^{\rm S}$ and the power of the driving voltage signal. The reproducibility and endurance of the resistive switching cycles benefit from the well-defined, few 10~nm scale, presumably single-domain volume of the IMT \cite{Posa2023}.

\textbf{Picosecond Time-Scale Set Switching.} The speed limit and energy consumption of resistive switching in VO$_{2}$ memristors was investigated by the real-time monitoring of the devices' resistive response to voltage pulses as short as 20~ps full width at half maximum (FWHM). Resistive switching taking place within 20~ps due to the injection of 1\,--\,5~fJ energy was demonstrated. The corresponding experimental setup is shown in Fig.~\ref{fig2}(c) and described in the Methods and Experimental Section. The measurement technique and the data analysis followed the procedures explained in great detail in Ref.~\citenum{Csontos2023}. In short, when the memristor device is exposed to fast voltage signals whose wavelength falls below the length of the utilized transmission lines, partial reflection and transmission of the $V_{\rm In}$ incoming voltage signal occurs due to the impedance mismatch between the $Z_{\rm MEM}$ device impedance and the $Z_{0}=$ 50~$\Omega$ wave impedance of the transmission lines. According to the solution of the telegraph equations applied to our experimental arrangement, the voltage drop on the sample equals to $2\cdot V_{\rm Refl}$ whereas the current equals to $V_{\rm Trans}/Z_{0}$, where $V_{\rm Refl}$ and $V_{\rm Trans}$ are the amplitudes of the reflected and transmitted harmonic waves, respectively. The impedance of the memristor can be determined through the formula
\begin{equation}
\frac{V_{\rm Trans}}{V_{\rm In}}=\frac{2Z_{0}}{Z_{\rm MEM}+2Z_{0}} \mbox{ ,}
\label{trans.eq}
\end{equation}
where, in general, harmonic $V_{\rm In}$ signals and a frequency dependent, complex-valued $Z_{\rm MEM}$ are assumed. When the memristor impedance is dominated by a frequency independent, real-valued resistive term, $Z_{\rm MEM}\approx R_{\rm MEM}$, Eq.~\ref{trans.eq} directly applies and $V_{\rm Trans}$ is proportional to any time dependent $V_{\rm In}$ signal. When a complex-valued, frequency dependent $Z_{\rm MEM}(f)$ is concerned, the numerical deduction of the device impedance requires a model assumption on $Z_{\rm MEM}(f)$ and a Fourier analysis based on Eq.~\ref{trans.eq}.

A simplistic equivalent circuit accounting for the $R_{\rm S}$ lead resistance and $C_{\rm MEM}$ parasitic capacitance of the memristor is shown in Fig.~\ref{fig2}(d). While the former contribution is usually negligible compared to the $R_{\rm MEM}$ device resistance, and merely plays a role in the capacitive response time through the $R_{\rm S}\cdot C_{\rm MEM}$ product, the frequency dependent capacitive impedance contribution arising from $C_{\rm MEM}$ may have a profound impact on the real time response during the applied $V_{\rm In}$ voltage pulses.

Figure~\ref{fig2}(a) shows the measured $V_{\rm Trans}$ response (blue, right axis) to a $V_{\rm In}$ pulse sequence (red, left axis). The latter consists of a 1~ns long, low-amplitude read-out pulse, a 20~ps FWHM, 1.3~V amplitude set pulse and a second read-out pulse identical to the first one. The relative timing of $V_{\rm In}$ and $V_{\rm trans}$ are compensated for the propagation time differences in the transmission lines according to the procedures outlined in Ref.~\citenum{Csontos2023}. During the first read-out pulse the device resides in its HRS and $V_{\rm Trans}$ stays low, consistently with Eq.~\ref{trans.eq} and $R_{\rm HRS}\approx$ 30~k$\Omega$. It is important to note the absence of a dominant capacitive peak in $V_{\rm Trans}$ during the 20~ps rise time of the first read-out pulse, exhibited in the magnified view of the inset in Fig.~\ref{fig2}(a). This observation evidences the negligible capacitive impedance of the device even in the HRS and indicates the purely resistive nature of the impedance switching. A quantitative analysis using LTspice shows a good agreement between the modeled (green) and measured (blue) $V_{\rm Trans}$ signal in the voltage step region by assuming $C_{\rm MEM}=$ 2~fF. In contrast, a higher $C_{\rm MEM}$ value would give rise to a dominant peak in response to the rising edge of $V_{\rm In}$, which is definitely not the case here.

Resistive switching due to the 20~ps FWHM set voltage pulse is demonstrated by the sharp increase of $V_{\rm Trans}$ during the set pulse and the persistence of an increased transmission during the second, 1~ns long read-out period. In the absence of a prevailing capacitive contribution, e.g., during the constant voltage read-out periods, the device's resistive impedance can be well approximated from the $V_{\rm Trans}/V_{\rm In}$ ratio based on Eq.~\ref{trans.eq}, as shown in Fig.~\ref{fig2}(b). Note, however, that the apparent rate of the such deduced resistance change within duration of the set pulse is instrumentally limited within the set pulse duration by the $\approx$60~GHz analog bandwidth of the detection setup. Additionally, within the duration of a short $V_{\rm In}$ pulse the frequency dependent capacitive contribution to $Z_{\rm MEM}$ can no longer be neglected which further limits the validity of the quantitative analysis on $R_{\rm MEM}(t)$ in this time window \cite{Csontos2023}.

\begin{figure*}[t!]
     \includegraphics[width=2\columnwidth]{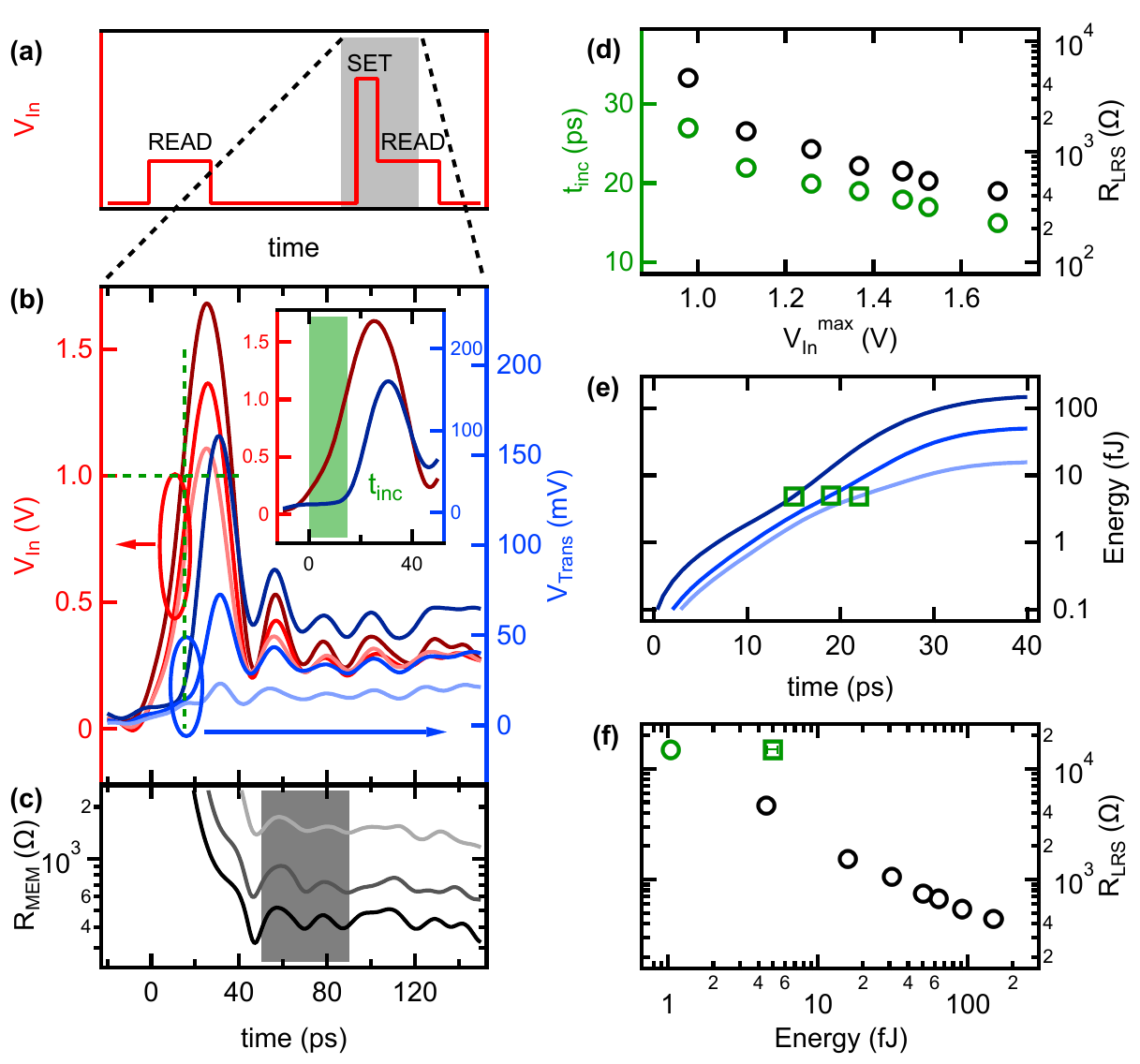}
     \caption{\textbf{Analysis of the set switching.} (a) Schematic illustration of the applied voltage pulse sequence. The first read pulse of 1~ns duration and 0.25~V amplitude is used to confirm the initial HRS. 100~ns later a 20~ps FWHM programming pulse of varying amplitude is applied which is directly followed by another 1~ns long, 0.25~V amplitude read pulse. Three $V_{\rm In}$ and $V_{\rm Trans}$ traces recorded in the shaded area are displayed in (b) in red (left axis) and blue colors (right axis), respectively. The horizontal green dashed line denotes $V_{\rm In}=$ 1~V whereas the corresponding vertical green dashed line highlights the onset of the switching due to the $V_{\rm In}^{\rm max}=$ 1.7~V pulse. The inset shows the magnified view of the highest amplitude traces of the main panel. The green shaded area illustrates the definition of the $t_{\rm inc}$ incubation time (see text for details). (c) Time dependent resistance calculated from the data presented in (b). The gray shaded area highlights the time window where averaging was applied to evaluate $R_{\rm LRS}$ after the completion of the set pulse. (d) Deduced $t_{\rm inc}$ (green, left axis) and $R_{\rm LRS}$ (black, right axis) as a function of the $V_{\rm In}^{\rm max}$ amplitude of the applied voltage pulses. (e) The solid lines show the set pulse energy calculated from the time dependent voltage and current for the three traces presented in (b). The green symbols mark the corresponding incubation times. (f) $R_{\rm LRS}$ as a function of the set pulse energy (black dots). The green square and its horizontal error bar correspond to the average and standard deviation of the energy integrals calculated until $t_{\rm inc}$, as shown in (e). The green circle is an estimated extrapolation of the energy dependence to $R_{\rm LRS}=$ 15~k$\Omega$.}
     \label{fig3}
\end{figure*}

Next, we investigate the set switching dynamics in more detail by applying 20~ps FWHM set voltage pulses of different amplitudes and evaluating the resulting resistance change, switching times and switching energies. The applied $V_{\rm In}$ pulse sequence is schematically illustrated in Fig.~\ref{fig3}(a). In essence, a similar pattern is utilized as discussed in the demonstrator experiment shown in Fig.~\ref{fig2}, only this time the first, 1~ns long and 0.25~V amplitude read-out pulse is shifted 100~ns away from the set pulse. As this time separation is much longer than the zero-bias relaxation time from the LRS to the HRS, a possible preconditioning of the set switching by the first read-out pulse can be unambiguously excluded.

Figure~\ref{fig3}(b) exemplifies the $V_{\rm Trans}$ response (blue colors, right axis) to 20~ps FWHM $V_{\rm In}$ pulses (red colors, left axis) of three different $V_{\rm In}^{\rm max}$ amplitudes. The corresponding, time dependent resistance traces assessed from the $V_{\rm In}/V_{\rm Trans}$ ratio according to Eq.~\ref{trans.eq} are shown in Fig.~\ref{fig3}(c). For further evaluation, the $R_{\rm LRS}$ values are deduced by averaging over the gray shaded time window in Fig.~\ref{fig3}(c). The fine analog tunability of the LRS is demonstrated in Fig.~\ref{fig3}(d) where a 60\% increase in $V_{\rm In}^{\rm max}$ results in the gradual, one order of magnitude decrease of $R_{\rm LRS}$ in the 10$^{2}$\,--\,10$^{3}$~$\Omega$ regime (black dots, right axis). This behavior is consistent with the set pulse energy dependent volume of the metal-insulator transition (MIT), as will be discussed later in the framework of our set switching model.

In addition to the decreasing tendency in $R_{\rm LRS}$, an apparent acceleration of the set transition is also observed at higher $V_{\rm In}^{\rm max}$. We characterize the speed of the set process by the $t_{\rm inc}$ incubation time which is experimentally defined here as the time interval between the 10\% onset of $V_{\rm In}^{\rm max}$ and reaching 15~k$\Omega\approx R_{\rm HRS}/2$, as illustrated by the green shaded area in the inset of Fig.~\ref{fig3}(b). The incubation time as a function of the applied $V_{\rm In}^{\rm max}$ setting is shown in Fig.~\ref{fig3}(d) by the green dots (left axis). Although the Mott type IMT in VO$_{2}$ is identified to have a dominant electronic origin and, thus, can be triggered even by femtosecond laser excitation \cite{Becker1994}, previous experiments utilizing voltage pulses have reported incubation times down to 100's of picoseconds \cite{Nikoo2022}. In contrast, we demonstrate incubation times down to 15~ps. Note, that the voltage pulse amplitude dependence does not show any saturation at this value, therefore $t_{\rm inc}$ values in the single digit picosecond regime are highly conceivable. However, the applied $V_{\rm In}^{\rm max}$ pulse amplitudes were restricted to $\leq$1.7~V in order to protect the devices from destruction.

\begin{figure*}[t!]
     \includegraphics[width=2\columnwidth]{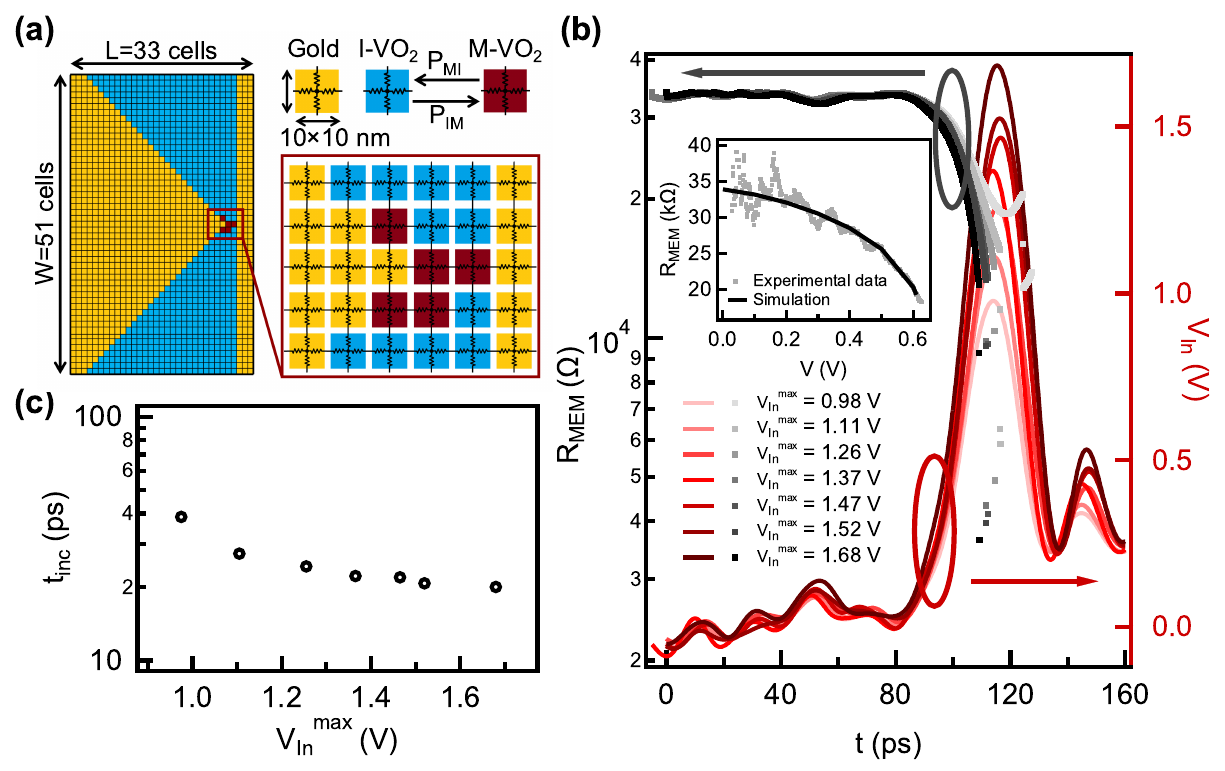}
     \caption{\textbf{Two-dimensional resistor network simulation of the set switching time.} (a) Illustration of the two-dimensional resistor network model applied for the experimentally realized electrode arrangement. The 40~nm thick VO$_{2}$ layer is in thermal contact with an electrically perfectly insulating substrate that is at T$_0$=$300\,$K. The yellow, blue and red squares represent low-resistivity Au, insulating phase VO$_{2}$ and metallic phase VO$_{2}$ domains, respectively. The dynamics of the transition between the latter two phases is determined by the transition probabilities $P_{\rm IM}$ and $P_{\rm MI}$. The width (length) of the simulated area is $W$=51 ($L$=33) cells. The gap between the electrodes at the narrowest spot is 2 cell wide. The resistance network connecting the nearest neighbor domains are also indicated. (b) Experimental $V_{\rm In}$ voltage pulses of 20~ps FWHM (red colors, right axis) and the corresponding, time-dependent calculated resistance traces (gray colors, left axis). The inset shows the $R(V)$ dependence in the HRS acquired in a DC measurement (gray dots) and its numerical fitting (black curve) which were used to extract the actual values of the thermal parameters of the model (see text). (c) Incubation times as a function of the $V_{\rm In}^{\rm max}$ set voltage pulse amplitudes, as deduced from the simulated data shown in (b).}
     \label{fig4}
\end{figure*}

The data presented in Fig.~\ref{fig3}(b) also sheds light on the threshold switching nature of the set transition: the lower $t_{\rm inc}$ values deduced at higher $V_{\rm In}^{\rm max}$ settings correspond to the same actual $V_{\rm In}\approx$1~V voltage levels, as highlighted by the green dashed lines. The observed apparent $V_{\rm In}^{\rm max}$ dependence of $t_{\rm inc}$ in Fig.~\ref{fig3}(d) is quantitatively accounted for by the instrumental aspects of our pulse firing setup, where the rise time of the $V_{\rm In}$ voltage pulse is always 20~ps, independently of the pulse amplitude. Consequently, the same threshold voltage is established faster when a higher amplitude pulse is applied. Such a threshold switching behavior is in agreement with the models which attribute a purely electronic \cite{Kim2004} or a mixed electronic and thermal origin \cite{Posa2023} to the IMT in VO$_{2}$. In contrast, ionic migration driven resistive switching compounds exhibit exponentially decreasing incubation times at linearly increasing voltage levels, known as the voltage-time dilemma \cite{Waser2009,Gubicza2015a,Santa2020}. Based on the above arguments, we argue that an independent control on $t_{\rm inc}$ and $R_{\rm LRS}$ shall be possible in VO$_{2}$ devices, as the former is merely governed by the rise time of the $V_{\rm In}$ set pulse while the latter is determined by its amplitude and duration, i.e., the total set pulse energy, as demonstrated by the black circles in Fig.~\ref{fig3}(d). Note, however, that $\leq$20~ps rise times at Volt-scale signal levels are facing the limitations of current state of the art electronics.

The set switching energies required to reach a specific $R_{\rm LRS}$ were calculated by numerically integrating the product of the $2\cdot V_{\rm Refl}$ voltage drop on the sample and the $V_{\rm Trans}/Z_{0}$ current as a function of time, as shown by the solid lines in Fig.~\ref{fig3}(e) for the three example time traces of Fig.~\ref{fig3}(b). The corresponding incubation times are marked by the green squares. They highlight that the set transition is ignited after the injection of $\approx$5~fJ energy, independently of the specific shape of the set pulse, underpinning the role of local heating in the IMT. The Joule heating contribution is estimated to be $\approx$1~fJ, by taking the capacitive charging energy of $C\cdot U^{2}/2\approx$ 4~fJ into account, where $U\approx2\cdot V_{\rm In}(t_{\rm inc})=$ 2~V and $C=$ 2~fF.

The LRS resistance values deduced from the 40~ps long time interval directly following the falling edges of the different amplitude set pulses are plotted against the total energy deposited in the device by the black circles in Fig.~\ref{fig3}(f). As a reference, the average and standard deviation of the total energies calculated at $t_{\rm inc}$ for the three example traces of Fig.~\ref{fig3}(e) are also displayed in Fig.~\ref{fig3}(f) by the green square and its error bar, respectively. According to our definition of $t_{\rm inc}$, the corresponding resistance value is 15~k$\Omega$. Note, that the energy values of the black circles naturally exclude most of the capacitive charging energy, as they involve the time integral over the entire duration of the corresponding set pulses where the contributions of charging and discharging mostly cancel out. Therefore, these energy values can be directly attributed to the Joule heating contribution. The green circle extrapolates the tendency drawn by the black symbols to $R_{\rm LRS}=$ 15~k$\Omega$ as an estimate of the set switching energy limitations for an ideally parasitic capacitance-free device design. In contrast, the energy value of the green square was determined at the onset of the set pulses, where the $\approx$ 4~fJ penalty of capacitive charging is not yet counteracted with the subsequent discharging. In comparison to the state of the art switching energies of 400~fJ in silicon CMOS neurons \cite{Cruz2012}, $\sim$50~fJ in electrochemical metalization cells \cite{Cheng2019} and $\sim$100~fJ in valence change oxide memories \cite{Strachan2011,Torrezan2011,Csontos2023} as well as in micrometer-scale VO$_{2}$ samples \cite{Nikoo2022}, this evaluation demonstrates the merits of nanoscale VO$_{2}$ devices in high-frequency electronics reaching single-digit femtojoule switching energies at k$\Omega$ range LRSs. The latter regime quantitatively corresponds to the extreme energy efficiency of the human brain, where the energy cost of a neural spike is estimated to be 5\,--\,100~fJ \cite{Sengupta2013,Yi2018}.

\begin{figure}[t!]
     \includegraphics[width=\columnwidth]{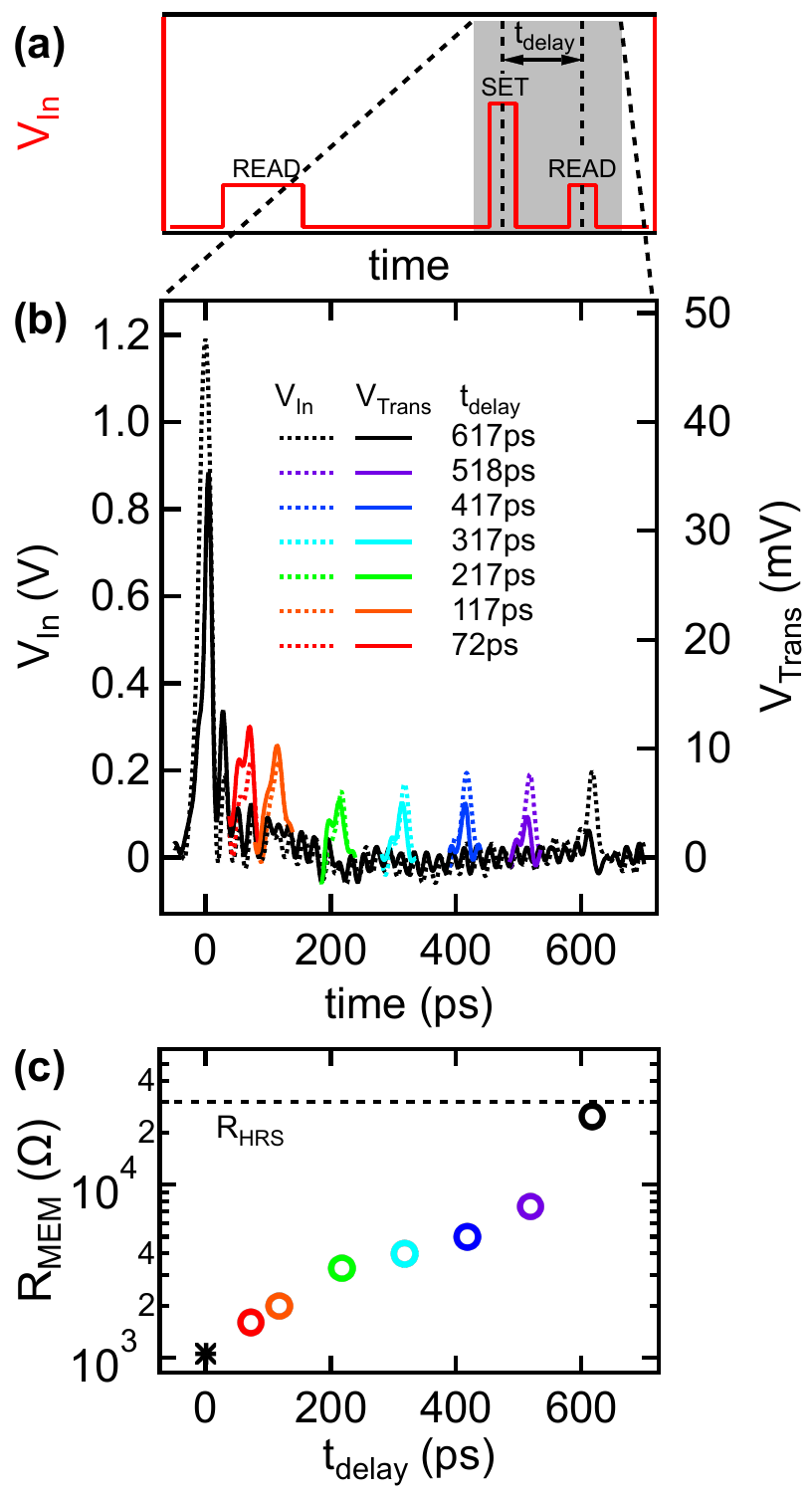}
     \caption{\textbf{Analysis of the retention.} (a) Schematic illustration of the applied voltage pulse sequence. Verification of the HRS and triggering the set transition are carried out identically to the scheme exhibited in Fig.~\ref{fig3}. In contrast, the second, 0.2~V amplitude read-out voltage pulse is only 20~ps long and is delayed with respect to the 20~ps long, 1.2~V amplitude programming pulse by a varying $t_{\rm delay}$. (b) $V_{\rm In}$ (dashed lines, left axis) and $V_{\rm Trans}$ data (solid lines, right axis) corresponding to the shaded area in (a) at different $t_{\rm delay}$ times, as labeled in the figure. Note, that only the second read-out pulse and its response are shown in color for better visibility. (c) Resistance as a function of the delay time, as deduced from the data shown in (b). The initial HRS resistance, indicated by the horizontal dashed line, is recovered within 600~ps.}
     \label{fig5}
\end{figure}

\textbf{Modeling the Set Switching Dynamics.} The plausibility of the set switching due to Volt-scale pulses of 20~ps FWHM within our thermally and electrically driven Mott transition picture \cite{Posa2023} was confirmed by numerical simulations. The applied two-dimensional resistor network model \cite{Stoliar2013,Rocco2022} takes into account the electrical and thermal conductivities of the VO$_{2}$ layer confined by the Au electrodes. The active region is modeled as an array of 10$\times$10~nm$^{2}$ cells arranged in a square lattice, as illustrated in Fig.~\ref{fig4}(a). The state of each VO$_{2}$ cell is either insulating (I-VO$_{2}$, blue) or metallic (M-VO$_{2}$, red) whereas the Au cells (yellow) are always in a low-resistance metallic state. The thickness of the cells is 40~nm according to the VO$_{2}$ film thickness of our samples. Each cell consists of four identical resistors connecting to its nearest neighbors. All VO$_{2}$ cells are initialized as I-VO$_{2}$.

In our model, the phase transition of the VO$_{2}$ cells is a thermally assisted process, where the $P_{\rm IM}(T)$ and $P_{\rm MI}(T)$ insulator-to-metal and metal-to-insulator transition probabilities depend exponentially on the temperature. The local temperature is determined from the heat equation accounting for Joule heating and heat conduction toward the neighboring cells and the substrate. Furthermore, a boundary thermal resistance was considered between the VO$_{2}$ and Au cells. The resistance of an individual resistor in an I-VO$_{2}$ cell depends on both the local temperature and the local electric field, whereas in the metallic state of VO$_{2}$ a temperature and electric field independent resistance is assumed. The electrical and thermal material parameter values utilized by the resistor network simulation are adopted from our previous finite element simulations performed in COMSOL Multiphysics \cite{Posa2023}. They were determined by fitting the experimental $R(V)$ trace of the device by the simulation, as shown in the inset of Fig.~\ref{fig4}(b). Further details on the numerical approach are provided in the Methods and Experimental Section as well as in Ref.~\citenum{Posa2023}.

In order to extract the incubations times according to the definition used throughout the evaluation of the measured set switching data, the experimentally realized set voltage pulses were applied to the resistor network, as shown by the red curves in Fig.~\ref{fig4}(b). The model was solved by using $0.1$~ps long time steps. The gray curves in Fig.~\ref{fig4}(b) show the time evolution of device resistance. The phase transition is hallmarked by the sudden drop of the resistance which occurs during the rise time of the set pulses, in agreement with the experiment. The simulated incubation times are exhibited in Fig.~\ref{fig4}(c). In spite of the crude simplifications of the applied model, they fall into the $20-40$~ps regime, in good agreement with the experimentally observed values of 15~ps$<t_{\rm inc}<$30~ps. The simulated incubation time also reveals a similar voltage dependence as its experimental counterpart, demonstrating the consistence of the measured, picosecond-scale incubation times with the electro-thermal picture of the Mott transition in VO$_{2}$ nanodevices. These findings are also in agreement with the Spice simulation results of W.~Yi et al. \cite{Yi2018}, where picosecond time-scale set transitions are predicted at femtojoule switching energies for active volumes comparable to our device design.

\textbf{Sub-Nanosecond Reset Dynamics.} Finally, we demonstrate that the HRS can be restored within 600~ps from the set voltage pulse. For this purpose, the pulsing scheme shown in Fig.~\ref{fig3}(a) is modified by replacing the second, 1~ns long read-out pulse promptly following the set pulse by a 20~ps long, 0.2~V amplitude probe pulse. The latter is delayed by $t_{\rm delay}$ with respect to the set pulse, as illustrated in Fig.~\ref{fig5}(a). The concept of the short probe pulse is introduced in order to minimize the impact of the readout pulse in maintaining the LRS. The initial $R_{\rm HRS}\approx$ 30~k$\Omega$ state is confirmed by a 1~ns long, 0.25~V amplitude read-out pulse applied 100~ns before the set pulse. The 20~ps long, 1.2~V amplitude set pulse, displayed by the black dashed line at 0~ps in Fig.~\ref{fig5}(b), switches the device into an $R_{\rm LRS}\approx$ 1~k$\Omega$ state within the duration of the pulse, consistently with the data exhibited in Fig.~\ref{fig3}. In the experiment shown in Fig.~\ref{fig5}(b), identical set pulses and different $t_{\rm delay}$ times between 0 and 600~ps were utilized, as plotted by the colored $V_{\rm In}$ (dashed lines) and $V_{\rm Trans}$ (solid lines) traces. The actual device resistance is evaluated at each $t_{\rm delay}$ by fitting the $V_{\rm Trans}$ response to the probe pulse according to Eq.~\ref{trans.eq} and the $Z_{\rm MEM}$ equivalent circuit shown in Fig.~\ref{fig2}(d). This analysis reveals the decay of the LRS toward the HRS and evidences that the latter can be restored within 600~ps, as presented in Fig.~\ref{fig5}(c).

Previous studies utilizing micron-scale VO$_{2}$ devices concluded that the MIT is a predominantly thermal relaxation driven, slow process taking place at 10's or even 100's of nanoseconds \cite{Valle2019}, where the presence of long-lived metallic domains was also evidenced \cite{Nikoo2022}. In contrast, our analysis unambiguously demonstrates that the relaxation to the HRS can take place at the sub-nanosecond time-scale in nanoscale VO$_{2}$ cells, as shown in Fig.~\ref{fig5}(c). This finding, together with the sub-100~ps IMT enables the orders of magnitude acceleration of VO$_{2}$-based electronics, among them THz sensors \cite{Qaderi2023} and oscillator circuits exploited for neuromorphic computing purposes \cite{Corti2020}.\\

\noindent \textbf{Conclusion}\\

In conclusion, we monitored the resistive switching dynamics in nanoscale volumes of VO$_{2}$ thin films in real time at the picosecond time-scale, enabled by a special, low-capacitance electrode design. By utilizing current state of the art electronics, we demonstrated that 20~ps FWHM voltage pulses of $\leq$1.7~V amplitude trigger the insulator-metal transition with incubation times down to 15~ps, resulting in fine-tunable, analog LRS  in the resistance range of 10$^{2}$\,--\,10$^{4}$~$\Omega$. These findings are supported by finite element simulations taking the combined electronic and thermal origin of the IMT into account. Our analysis of the energy consumption revealed that, depending on the targeted $R_{\rm LRS}$ level, the set switching requires as little energy as $\approx$4~fJ. By applying a pump-probe pulsing scheme we demonstrated that the HRS can be restored within 600~ps. The above results represent orders of magnitude breakthroughs both in the operation frequency and energy efficiency, demonstrating the merits of nanoscale Mott devices in the electronic platforms of the Beyond-Moore Era.\\

\noindent \textbf{Methods and Experimental Section}\\

\textbf{Device Fabrication.} The VO$_{2}$ layers were formed via the post-deposition heat treatment of an Al$_{2}$O$_{3}$/V vertical stack of 480~$\mu$m Al$_{2}$O$_{3}$ and 100~nm V. During the heat treatment, the sample was exposed to 400~$^{\circ}$C temperature and 0.1~mbar air pressure over 4.5~hours. As a result, a 40~nm thick VO$_{2}$ film was created on top of a 180~nm thick V$_{2}$O$_{5}$ bottom layer, as confirmed by cross-sectional TEM and EELS analyses \cite{Posa2023}. The metallic leads of 10~nm Ti and 50~nm Au were patterned by standard electron-beam lithography and deposited by electron-beam evaporation at 10$^{-7}$~mbar base pressure at rates of 0.1~nm/s and 0.4~nm/s, respectively, followed by lift-off. After the completion of the CPW structure shown in Fig.~\ref{fig1}(c), a selective etching step was carried out to remove the VO$_{2}$ layer in the gap regions of the GND-S and GND-D electrodes of the CPW. During this step the sample was immersed to an acidic solution of H$_{2}$O$_{2}$:H$_{3}$PO$_{4}$:CH$_{3}$COOH:HNO$_{3}$ (2:16:1:1) at 50$^{\circ}$C for twice 5 seconds. This method yielded to $\sim$10~M$\Omega$ parasitic resistances between the GND-S and GND-D electrodes of the CPW.

\textbf{Direct Current (DC) Characterization.} The schematic of the DC $I(V)$ measurement is shown in Fig.~\ref{fig1}(d). A slow, typically $f_{\rm Drive}$=1~Hz frequency, triangular voltage signal was applied to the device under test and the series resistor of $R^{\rm S}$ = 0.3\,--\,1~k$\Omega$ by an NI USB-6341 data acquisition unit (DAQ). The current was measured by a Femto DHPCA-100 current amplifier and recorded at the analog voltage input of the DAQ. The $V_{\rm Bias}$ voltage acting on the device was determined as $V_{\rm Bias}=V_{\rm Drive}-I\cdot R^{\rm S}$.

\textbf{Fast Switching Setup.} The schematics of the fast resistive switching setup is shown in Fig.~\ref{fig2}(c). The device under test was contacted by two 67~GHz bandwidth Picoprobe triple probes in a vibration-damped probe station. A Micram DAC10004 100GSa/s DAC unit served as an arbitrary waveform generator (AWG) which provided voltage pulses down to 20~ps FWHM at 20~ps rise time. The output of the AWG was amplified by a Centellax UA0L65VM broadband amplifier module owing a 65~GHz analog bandwidth. The voltage pulses propagated in 0.30~m long, 70~GHz bandwidth, 50~$\Omega$ terminated Totoku TCF280 coaxial cables. The $V_{\rm Trans}$ transmitted voltage was recorded by a 50~$\Omega$ terminated Keysight UXR1104A digital storage oscilloscope (DSO) at 256~GSa/s sampling rate and 113~GHz analog bandwidth. The input terminals of the DSO were protected by 60~GHz bandwidth RF attenuators. The $V_{\rm In}$ signal was acquired separately by eliminating the memristor device and the probes from the circuit.

\textbf{Two-Dimensional Resistor Network Simulations.} The $R_{I}^{i,j}$ resistance of an individual resistor in the I-VO$_{2}$ cell, indexed by ($i,j$), depends on the $T_{i,j}$ local temperature and $E_{i,j}$ electric field and can be written as
\begin{equation}
R_{I}^{i,j}= R_{0}e^{-\frac{E_{g}}{2k_{B}T_{i,j}}}e^{\frac{E_{i,j}}{E_{c}}} \mbox{,}
\end{equation}
where $R_{0}$ is a constant, $E_{g}$ is the band gap energy in the HRS state and $E_{c}$ is a characteristic electric field. In the LRS a temperature and electric field independent resistance value $R_{M}^{ij}$ is assumed. By biasing the resistor network, the current starts to flow through the cells and the Joule heat dissipates on each resistor according $P_{i,j}=I_{i,j}^2R_{I}^{i,j}$. The corresponding $dT_{i,j}$ temperature change depends on the Joule heating as well as on the heat conduction towards the nearest neighbor cells and the substrate according to
\begin{equation}
\frac{dT_{i,j}}{dt} = \frac{P_{i,j}}{C_{i,j}}-\frac{\kappa_{i,j}}{C_{i,j}} \sum_{k,l}^{NN}(T_{i,j}-T_{k,l}) -\frac{\kappa_{\rm subs}}{C_{i,j}}(T_{i,j}-T_{0}) \mbox{,}
\label{heat.eq}
\end{equation}
where $C_{i,j}$ is the thermal capacity of the cell, $\kappa_{i,j}$ ($\kappa_{\rm subs}$) is the thermal conductance toward the neighbor cells (substrate). The substrate is assumed to be at a $T_{0}$ base temperature. Furthermore, a boundary thermal resistance was considered between the VO$_{2}$ and Au cells which determines the $\kappa_{\rm int}$ thermal conductance at their interface. At each time step of the simulation, the resistance of the resistor network is calculated and the electrical potential map of the resistor network is determined by a nodal analysis. Finally, $T_{i,j}$ is calculated via Eq.~\ref{heat.eq} and the state of each VO$_{2}$ cell is updated.

The phase of a VO$_{2}$ cell can change between metallic and insulating states, according to the transition probabilities given by $P_{\rm IM}$ and $P_{\rm MI}$. These transitions are thermally activated with the transition rates of 
\begin{eqnarray}
P_{\rm IM}&=&\nu e^{-\frac{E_{\rm IM}(T)}{k_{B}T}}\\
P_{\rm MI}&=&\nu e^{-\frac{E_{\rm MI}(T)}{k_{B}T}} \mbox{,}
\end{eqnarray}
where the $\nu$ is an attempt rate set to unity. The potential barriers $E_{\rm IM}(T)$ and $E_{\rm MI}(T)$ separate the metal and insulator states. Both energy barriers have a linear temperature dependence, which vanishes at the phase transition temperature $T_{\rm c,heat}$ and $T_{\rm c,cool}$, respectively, as
\begin{equation}
E_{\rm IM}(T)= 
\begin{cases}
    \epsilon_{\rm IM}(T_{\rm c,heat}-T),& \text{if } T < T_{\rm c,heat}\\
    0,              & \text{if } T\geq T_{\rm c,heat}
\end{cases}
\end{equation}
\begin{equation}
E_{\rm MI}(T)= 
\begin{cases}
    \epsilon_{\rm MI}(T-T_{\rm c,cool}),& \text{if } T > T_{\rm c,cool}\\
    0,              & \text{if } T\leq T_{\rm c,cool}
\end{cases}
\end{equation}
where $\epsilon_{\rm IM}$ and $\epsilon_{\rm MI}$ are constants. The electrical parameters $R_{0}$ and $E_{c}$ were deduced from the measured low-bias $R(V)$ dependence. The thermal parameters $\kappa_{\rm VO2}$, $\kappa_{\rm subs}$ and $\kappa_{\rm int}$ were determined by fitting the high-bias $R(V)$ trace in the insulating phase.\\

\noindent \textbf{Author Contributions}\\

T.N.T. and L.P. fabricated the devices. The VO$_{2}$ layers were grown by G.M.  M.C and Y.H. developed the fast resistive switching setup. S.W.S. and M.C. acquired and analyzed the resistive switching data. B.S. and Z.P. contributed to the lower frequency characterization of the samples. L.P. developed the two-dimensional resistor network model and performed the numerical simulations. A.H. and M.C. conceived the idea of fast resistive switching experiments and supervised the project. The manuscript was prepared by M.C., S.W.S, L.P. and A.H. All authors contributed to the discussion of the results.\\

\noindent \textbf{Acknowledgements}\\

This research was supported by the Ministry of Culture and Innovation and the National Research, Development and Innovation Office within the Quantum Information National Laboratory of Hungary (Grant No. 2022-2.1.1-NL-2022-00004), and the NKFI K143169 and K143282 grants. Project no. 963575 has been implemented with the support provided by the Ministry of Culture and Innovation of Hungary from the National Research, Development and Innovation Fund, financed under the KDP-2020 funding scheme. L.P. acknowledges the support of the UNKP-23-5-BME-422 new national excellence program of the Ministry for Innovation and Technology from the source of the National Research, Development and Innovation Fund and the J\'{a}nos Bolyai Research Scholarship. J.L. and M.C. acknowledge the financial support of the Werner Siemens Stiftung.\\

\bibliography{References}

\end{document}